# Negative strange quark-chemical potential
# A necessary and sufficient observable of the
# deconfinement phase transition in a finite-baryon equilibrated state*

*Apostolos D. Panagiotou and Panayiotis Katsas*

*University of Athens, Physics Department*
*Nuclear & Particle Physics Division*
*Panepistimiopolis, GR-157 71 Athens, Hellas*

**Abstract**

We consider the variation of the strange quark-chemical potential in the phase diagramme of nuclear matter, employing the order parameters and mass-scaled partition functions in each domain and enforcing flavour conservation. Assuming the region beyond the hadronic phase to be described by massive, correlated and interacting quarks, in the spirit of lattice and N−J-L calculations, we find the strange quark-chemical potential to attain large negative values in this domain. We propose that this change in the sign of the strange quark-chemical potential, from positive in the hadronic phase to negative in the partonic, to be a unique, concise and well-defined indication of the quark-deconfinement phase transition in nuclear matter. We propose also that, for a chemically equilibrated state in the deconfined region, which follows an isentropic expansion to hadronization via a second order phase transition, the fugacities of the equilibrated quark flavours, once fixed in the primordial state, remain constant throughout the hadronization process.

e-mail: apanagio@phys.uoa.gr

\*   Work supported in part by the Research Secretariat of the University of Athens.

# 1. Introduction

It is generally expected that ultra-relativistic nucleus-nucleus collisions will provide the basis for strong interaction thermodynamics, which will lead to new physics. Quantum Chromodynamics for massless quarks free of dimensional scales contain the intrinsic potential for the spontaneous generation of two scales: one for the confinement force, coupling quarks to form hadrons and one for the chiral force, binding the collective excitations to Goldstone bosons. In thermodynamics, these two scales lead to two possible consecutive phase transitions, deconfinement and chiral symmetry restoration, characterized by corresponding critical temperatures: $T_d$ and $T_\chi$, [1]. At temperatures above $T_d$, hadrons dissolve into quarks and gluons, whereas at $T_\chi$, chiral symmetry is fully restored and quarks become massless (current-mass), forming the ideal Quark - Gluon Plasma (QGP). A priori it is not evident that both non-pertubative transitions have to take place at the same temperature. At finite net baryon number density, $T_d < T_\chi$ would correspond to a regime of unbound, massive, correlated and interacting, 'constituent-like' quarks [2], as they appear in the additive quark model for hadron-hadron and hadron-lepton interactions [3]. Thus, the consecutive appearance of these two transitions, deconfinement and chiral symmetry restoration, forms an intermediate region on the phase diagramme, the Deconfined Quark Matter (DQM), in-between the Hadron Gas (HG) and the ideal QGP domains. Therefore, we define a QCD phase diagramme of strongly interacting matter at finite baryon number density with three regions: HG - DQM - QGP [4-6].

The necessity for such an intermediate region with massive quarks and gluons is conjectured also from recent lattice calculations [7], which show that $\varepsilon(2T_d) \sim 0.85\varepsilon_{SB}$ and $3P(2T_d) \sim 0.66\varepsilon_{SB}$, where $\varepsilon_{SB}$ is the Stefan-Boltzmann ideal QGP value for the energy density. In addition, the running coupling constant $\alpha_s(T\sim 300$ MeV$) \sim 0.3$. These observations substantiate quantitatively the need for an intermediate domain with $1 > \alpha_s > 0$, where massive and correlated- interacting quarks are found. This domain does not have a defined upper border, but goes asymptotically into the ideal QGP region with increasing temperature and quark-chemical potential.

Such a 3-state phase diagramme could be described by the variation of thermodynamic quantities from one region to the other. The task is to establish a well-defined quantity, which changes concisely (and measurably) as nuclear matter changes phases, thus indicating deconfinement and / or chiral symmetry restoration. If this quantity can be expressed in functional form of other thermodynamic parameters within an Equation of State (EoS), one may then describe its variation throughout the phase diagramme. We propose and show that the equilibrated strange quark-chemical potential of the strongly interacting system is the sought-for thermodynamic quantity. A similar suggestion was put forth earlier [5], but confident experimental data to substantiate it were not available at the time.

In sections 2, 3 and 4 we discuss the partition functions in the HG, ideal QGP and DQM phases for the strange hadron sector, respectively. Taking into account in an approximate way the dynamics of the DQM phase, we construct an empirical Equation of State for this domain. We employ it, together with the known EoS of the HG and QGP phases, to obtain the strange quark-chemical potentials in functional form of the variables temperature and light quark-chemical potential throughout the phase diagramme, $\mu_s = f(T,\mu_q)$. With this relation we study the variation of $\mu_s$ in the 3-region phase diagramme, attributing the changes in the sign and magnitude of the strange quark-chemical potentials to the changes of phase of nuclear matter.

In section 5 we summarize several thermal model analyses of experimental particle yield data from nucleus-nucleus interactions at AGS and SPS. Among the latter, the S+A interactions at 200 AGeV indicate the existence of negative strange quark-chemical potential. We compare these results with our proposal. In section 6 we predict the values for certain strange particle ratios and suggest experiments at lower energies at RHIC, where the baryon number density is finite. Finally, in section 7 we discuss our proposals and come to conclusions whether the S+A and Pb+Pb interactions at the SPS have entered the deconfined phase.

**2. Hadron Gas phase**

In the HG phase, the mass spectrum is given by the partition function, $\ln Z_{HG}$, in the Boltzmann approximation. We assume that the hadronic state has attained thermal and chemical equilibration of three quark flavors (u,d,s):

$$\ln Z_{HG}(T, V, \lambda_q, \lambda_s) = Z_m + Z_n(\lambda_q^3 + \lambda_q^{-3}) + Z_K(\lambda_q \lambda_s^{-1} + \lambda_s \lambda_q^{-1}) + Z_Y(\lambda_s \lambda_q^2 + \lambda_q^{-2}\lambda_s^{-1})$$
$$+ Z_\Xi(\lambda_s^2 \lambda_q + \lambda_s^{-2}\lambda_q^{-1}) + Z_\Omega(\lambda_s^3 + \lambda_s^{-3}) \qquad (q = u, d) \qquad (1)$$

The one particle Boltzmann partition function $Z_k$ is given by

$$Z_k(V, T) = \frac{VT^3}{2\pi^2} \sum_j g_j \left(\frac{m_j}{T}\right)^2 K_2\left(\frac{m_j}{T}\right) \qquad (2)$$

The fugacity $\lambda_i^k$ controls the quark content of the k-hadron; i = s, b for strangeness and baryon number, $\lambda_b = \lambda_q^3 = \exp(3\mu_q/T)$, respectively. The summation in Eq. (2) runs over the resonances of each hadron species with mass $m_j$, and the degeneracy factor $g_j$ counts the spin and isospin degrees of freedom of the j-resonance. For the strange hadron sector, kaons with masses up to 2045 MeV/$c^2$, hyperons up to 2350 MeV/$c^2$ and cascades up to 2025 MeV/$c^2$ are included, as well as the known $\Omega^-$ states at 1672 MeV and 2252 MeV. For simplicity, we assume isospin symmetry in Eq. (1), $\mu_u = \mu_d = \mu_q$.

Strangeness neutrality in strong interactions necessitates:

$$< N_s - N_{\bar{s}} > = \frac{T}{V}\frac{\partial}{\partial \mu_s}\left[\ln Z_{HG}(V,T,\lambda_s,\lambda_q)\right] = 0 \qquad (3)$$

which reduces to

$$Z_K(\lambda_s\lambda_q^{-1} - \lambda_q\lambda_s^{-1}) + Z_Y(\lambda_s\lambda_q^2 - \lambda_q^{-2}\lambda_s^{-1}) + 2Z_\Xi(\lambda_s^2\lambda_q - \lambda_s^{-2}\lambda_q^{-1}) + 3Z_\Omega(\lambda_s^3 - \lambda_s^{-3}) = 0 \qquad (4)$$

This is an important condition as it defines the relation between light- and strange-quark fugacities in the equilibrated primordial state. Eq. (4) can be used to derive the *true* (transverse-flow independent) temperature of the state, once the fugacities $\lambda_q$ and $\lambda_s$ are known from experimental strange particle yield ratios [4].

In the HG phase with finite net baryon number density, the chemical potentials $\mu_q$ and $\mu_s$ are coupled through the production of strange hadrons. Due to this coupling, strangeness conservation does not necessitate $\mu_s = 0$ everywhere in this phase. In fact, $\mu_s > 0$ in the hadronic domain. The condition $\mu_s = 0$ requires $\lambda_s = \lambda_s^{-1} = 1$, and Eq. (4) becomes [8],

$$[Z_Y(\lambda_q + \lambda_q^{-1}) - Z_K + 2Z_\Xi](\lambda_q - \lambda_q^{-1}) = 0$$

The first factor gives the curve $\mu_q$ as a function of T in the phase diagramme along which $\mu_s = 0$:

$$\mu_q(T) = T\cosh^{-1}(Z_K/2Z_Y - Z_\Xi/Z_Y) \qquad (5)$$

A more elegant and concise formalism describing the HG phase is the Strangeness-including Statistical Bootstrap model (SSBM) [9]. It includes the hadronic interactions through the mass spectrum of all hadron species, in contrast to all other ideal hadron gas formalisms. The SSBM is valid and applicable only within the hadronic phase, defining in a determined way the limits of this phase. The boundary of the hadronic domain is given by the projection on the 2-dimentional $(T,\mu_q)$ phase diagramme of the intersection of the 3-dimentional bootstrap surface with the strangeness-neutrality surface ($\mu_s = 0$). Note that the vanishing of the strange quark-chemical potential on the HG borderline does not *apriori* suggest that $\mu_s = 0$ everywhere beyond. It only states that this zero value for the strange quark-chemical potential characterizes the end of the hadronic phase.

### 3. Ideal QGP Phase

In the ideal QGP region, the EoS for the current-mass u,d,s-quarks and gluons has the form:

$$\ln Z_{QGP}(T,V,\mu_q,\mu_s) = \frac{V}{T}\left\{\frac{37}{90}\pi^2 T^4 + \mu_q^2 T^2 + \frac{\mu_q^4}{2\pi^2} + \frac{g_s m_s^{0\,2} T^2}{2\pi^2}\left(\lambda_s + \lambda_s^{-1}\right)K_2\left(\frac{m_s^0}{T}\right)\right\} \quad (6)$$

where $m_s^o$ and $g_s$ is the current-mass and degeneracy of the s-quark, respectively. Strangeness conservation gives $\lambda_s = \lambda_s^{-1} = 1$, or

$$\mu_s^{QGP}(T,\mu_q) = 0 \quad (7)$$

throughout this phase. In this region the two order parameters, the average thermal Wilson loop $<L>$ and the scalar quark density $<\bar{\psi}\psi>$, have reached their asymptotic values.

### 4. DQM Phase

In formulating our description beyond the hadronic phase, we use the following picture: The thermally and chemically equilibrated primordial state with finite baryon number density, produced in nucleus-nucleus interactions, consists of the deconfined valance quarks of the participant nucleons, as well as of q-$\bar{q}$ pairs (q=u,d,s), created by quark and gluon interactions. Beyond but near the HG boundary, $T \geq T_d$, the correlation-interaction between q-q is near maximum, $\alpha_s(T) \leq 1$, a prelude to confinement into hadrons upon hadronization. With increasing temperature, the correlation-interaction of the deconfined quarks gradually weakens, $\alpha_s(T) \to 0$, as colour mobility and colour charge screening increase. The masses of all (anti)quarks depend on the temperature of the state and scale according to a prescribed way. The initially constituent mass decreases and as the DQM region goes asymptotically into the ideal QGP domain, as $T \to T_\chi$, quarks attain current-mass. In this formulation, the equation of state in the DQM region should lead to the EoS of the hadronic phase, Eq. (1), at $T < T_d$ and to the EoS of the ideal QGP, Eq.(6), at $T \sim T_\chi$.

To construct the empirical EoS in the DQM phase, we use the two order parameters (for details see Appendix A):

(a)  The average thermal Wilson loop, $<L> = \exp(-F_q/T) \sim R_d(T) = 0 \to 1$, as $T = T_d \to T_\chi$, describing the quark deconfinement and subsequent colour mobility, $F_q$ being the free quark energy.

(b)  The scalar quark density, $<\bar{\psi}\psi> \sim R_\chi(T) = 1 \to 0$, as $T = T_d \to T_\chi$, denoting the scaling of the quark mass with temperature.

We assume that above $T_d$ the deconfined quarks retain a degree of correlation, resembling "hadron-like" states, since $1 > \alpha_s > 0$. The diminishing of this correlation-interaction, as a result of progressive increase of colour mobility, is approximated by the factor: $[1-R_d(T)] = 1 \to 0$, as $T = T_d \to T_\chi$. Note that effectively $[1-R_d(T)] \sim \alpha_s(T)$ in the DQM region, the temperature dependence of the running coupling constant [10]. To account for the ''effective mass'' of the state as a function of temperature, we assume the mass of the quarks to decrease from the constituent value and reach the current-mass as $T \to T_\chi$. The quark mass scales with temperature as:

$$m_q^*(T) = R_\chi(T)(m_q - m_q^o) + m_q^o, \qquad (8)$$

where $m_q$ and $m_q^o$ are the constituent and current quark masses, respectively, ($m_u^o \sim 5$ MeV, $m_d^o \sim 9$ MeV, $m_s^o \sim 170$ MeV). Similarly, the 'effective hadron' mass scales as:

$$m_i^*(T) = R_\chi(T)(m_i - m_i^o) + m_i^o, \qquad (9)$$

where $m_i$ is the hadron mass in the hadronic phase and $m_i^o$ is equal to the sum of the hadron's quarks current-mass ($m_K^o \sim 175$ MeV, $m_Y^o \sim 185$ MeV, $m_\Xi^o \sim 350$ MeV and $m_\Omega^o \sim 510$ MeV). In the EOS, the former scaling is employed in the mass-scaled partition function $\ln Z_{QGP}^*$, whereas the latter in the mass-scaled partition function $\ln Z_{HG}^*$, which accounts for the produced hadron species. Note that this mass-scaling is effectively equivalent to the one given in the Nambu – Jona-Lasinio (N–J-L) formalism [11] (see Appendix A).

Employing the described dynamics, we construct the empirical EoS of the DQM phase:

$$\ln Z_{DQM}(V,T,\lambda_q,\lambda_s) = [1 - R_d(T)]\ln Z_{HG}^*(V,T,\lambda_q,\lambda_s) + R_d(T)\ln Z_{QGP}^*(V,T,\mu_q,\mu_s) \qquad (10)$$

The factor $[1-R_d(T)]$ describes the weakening of the correlation-interaction of the deconfined quarks constituting the "hadron-like" entities and $\ln Z_{HG}^*$ gives the mass-scaling of these entities with increasing temperature. In the second term, the factor $R_d(T)$ defines the rate of colour mobility, whereas $\ln Z_{QGP}^*$ represents the state as it approaches the QGP region. Thus, at $T = T_d$, the EoS of the DQM region goes over to the corresponding in the HG phase and at $T \sim T_\chi$, to the EoS in the ideal QGP region. In this calculation we have taken $T_\chi = 3.5 T_d$.

Strangeness neutrality in the DQM phase leads to:

$$[1 - R_d(T)][Z_K^*(\lambda_s\lambda_q^{-1} - \lambda_q\lambda_s^{-1}) + Z_Y^*(\lambda_s\lambda_q^2 - \lambda_s^{-1}\lambda_q^{-2}) + 2Z_\Xi^*(\lambda_s^2\lambda_q - \lambda_s^{-2}\lambda_q^{-1})$$
$$+ 3Z_\Omega^*(\lambda_s^3 - \lambda_s^{-3})] + R_d(T)g_s m_s^{*2} K_2\left(\frac{m_s^*}{T}\right)(\lambda_s - \lambda_s^{-1}) = 0 \qquad (11)$$

For given $\mu_q$, Eq. (11) defines the variation of the strange quark-chemical potential with temperature in the DQM domain. Combining Eq's (4,11) we obtain the variation of the strange quark-chemical potential with temperature in the entire phase diagramme.

Fig. 1 maps the QCD phase diagramme as a function of T and the strange quark-chemical potential. It exhibits the behaviour of $\mu_s$ with temperature for fixed $\mu_q = 0.45T$. The strange quark-chemical potential attains positive values in the HG phase, as a result of the coupling between $\mu_q$ and $\mu_s$ in strange hadrons. It approaches zero as the hadron density reaches its asymptotic Hagedorn limit [12] at the end of the hadronic phase, where a phase transition to partonic matter takes place, at the deconfinement temperature $T_d$. At this temperature, $\mu_s(T_d) = 0$, signifying the vanishing of this coupling in hadrons. In the DQM region it grows strongly negative, where an effective coupling − progressively weakening − remains in effect among the deconfined, but correlated and interacting quarks with non-current mass. Finally $\mu_s$ returns to zero as the ideal QGP phase is approached with $\alpha_s \sim 0$ and current-mass quarks.

Fig. 2 exhibits the variation of $\mu_s/T$ as a function of T throughout the 3-region phase diagramme for $\mu_q = 0.45T$. We have assumed that the deconfinement temperature is $T_d \sim 176$ MeV at $\mu_q \sim 80$ MeV, as given by the SSBM hadronic phase boundary [18]. We observe that:

α) If upon deconfinement quarks attain instantaneously current mass and $\alpha_s(T>T_d) = 0$, the ideal QGP phase follows immediately after the HG phase and the EoS beyond the HG domain is given by Eq. (6). In this case $\mu_s(T) = 0$ for all $T > T_d$.

b) If $0<\alpha_s(T>T_d)<1$ and $m_q^0<m_q^*(T>T_d)<m_q$, then the EoS (10), which includes mass-scaled "hadron-like" states and variable $\alpha_s(T) \sim 1 - R_d(T)$, as well as (anti)quarks with finite (scaled) mass, gives large negative values for $\mu_s/T$ in the DQM phase, which, after reaching a minimum, returns asymptotically to zero as the ideal QGP phase is approached.

c)      Scaling alone of the effective hadron masses produces large negative strange quark-chemical potential beyond the HG phase, saturating at high temperature without ever returning to zero at the ideal QGP phase.

On the basis of the above observations, we propose that the change of sign of the strange quark-chemical potential from positive to negative defines uniquely and precisely the phase transition to quark-deconfinement. We note that the sign of $\mu_s$: positive in the HG, negative in the DQM and zero in the ideal QGP domains, is independent of the particular form of the parameters $R_d$ and $R_\chi$ used in the EoS and unique in each region (see Appendix A). It is also independent of assumptions and ambiguities regarding interaction mechanisms and matter media effects, as well as weakness and uncertainties of models, as is the case with other proposed signatures for deconfinement: $J/\psi$ suppression, resonance shift and broadening, strangeness enhancement, etc. In the present calculation, however, the magnitude of the negative chemical potential should be taken in a qualitative manner, due to the empirical treatment of the dynamics in the DQM phase. A detailed, quantitative treatment of the EoS in the DQM phase will require the use of a three-flavour effective Lagrangian in the N−J-L formalism [13].

## 5.    Experimental Data

Data from nucleus-nucleus interactions, obtained by experiments E802 [14] at AGS and NA35 [15], NA49 [16] at SPS have been analyzed in terms of several statistical-thermal models: the Strangeness-including Statistical Bootstrap Model, SSBM [9,17-20] and others employing the canonical and grand-canonical formalisms [21-25]. Table 1 summarizes the results of these analyses, from which the quantities T, $\mu_q$ and $\mu_s$ have been deduced. Fig. 3 is a plot of the mean values of the temperature and strange quark-chemical potential, obtained from these calculations. Observe that the interactions Si+Au, Au+Au and Pb+Pb at both energies have positive $\mu_s$, whereas the interactions S+S and S+Ag at 200 AGeV exhibit negative values. This is the first confident [5] experimental confirmation of negative values for the strange-quark-chemical potential. It should be noted that earlier thermal model fits [23] to preliminary Pb+Pb data of NA49 [26], with larger $\varphi$ and $\Xi$ particle yields than the final values [16], produced a high temperature of 193 MeV and $\mu_s = -72$ MeV for the equilibrated state. These values, although wrong, show that the strange quark-chemical potential could attain even larger negative values at higher temperatures, a systematic trend.

In contrast to other thermal-statistical models, the Statistical Bootstrap Model (SBM) of Hagedorn [27] incorporates the hadronic interactions in the EoS through the bootstrap equation and thus gives, in a definitive way, the limits of the hadronic phase as a result of the branch point of this equation. The development of the SBM with the inclusion of Strangeness (SSBM) [9,17-19] has been employed in the analysis of nucleus-nucleus interactions at the SPS. The SSBM analysis for the S+S interaction [18] has shown that this equilibrated state is situated mostly (75%) outside the hadronic phase, whereas in the case of S+Ag [19], it is just on the deconfinement line. In addition, the calculations have pointed out a large (~30%) entropy enhancement of the experimental data compared to the model for both systems, an effect observed also by other calculations [21-23]. This enhancement may be attributed to contributions from the DQM phase with many liberated new partonic degrees of freedom. The fact that beyond the HG phase the deconfined quarks retain a degree of correlation resembling "hadron-like" states, allows the thermal models [21-23] to describe the state adequately using a hadron gas EoS. These models do not sense the HG phase limit and treat the deconfined state as consisting of colourless hadrons. This may not lead to erroneous results for a domain in the phase diagramme not too far away from the deconfinement line.

For the Pb+Pb interaction, the SSBM analysis [20] has shown the system to be located well within the hadronic phase. In addition, thermal model calculations [20,24] find no entropy enhancement for this interaction. These results are corroborating the observation of positive strange quark-chemical potential for the Pb+Pb and negative for the S+A interactions, positioning the former within and the latter beyond the HG phase, in the DQM domain. Fig. 4 shows the phase diagramme with the SSBM maximally extended[1] deconfinement line and the location (T, $\mu_q$ mean value) of several interactions, Table 1. Observe that the sulfur-induced interactions at 200 AGeV, having negative $\mu_s$, are situated beyond the deconfinement line, whereas all others with positive $\mu_s$ are located well within the hadronic phase.

## 6. Predictions for particle yield ratios at RHIC

New data obtained at RHIC with the Au+Au interaction at $\sqrt{s}$ = 130 AGeV (see a compilation of data in [28]) show that at midrapidity the light quark fugacity is $\lambda_q$ = 1.08, indicating very small quark-chemical potential. This has an effect on the strange quark-chemical potential, making it very small and close to zero throughout the 3-region phase diagramme. In such case of minimal net baryon density, our model cannot be applied. We have, therefore, calculated the particle yield ratios $\Omega^+/\Omega^-$ and $\Xi^+/\Xi^-$ for a finite baryon density region, obtained in Au+Au interactions at $\sqrt{s}$ = 20 - 90 AGeV. The quantities needed for these calculations ($\mu_q$/T and T) were obtained by fitting all available corresponding values, obtained for equilibrated states in nucleus-nucleus collisions from SIS to SPS energies [24,25] and extrapolating to RHIC energies. The data were fitted as a function of ($\sqrt{s}$/participant). Table 2 contains the extrapolated values for T and $\mu_q$/T, corresponding to equilibrated primordial states (not chemical freeze out), as well as predictions of our empirical EoS for $\mu_s$/T and for the two particle ratios. Fig. 5 shows the predicted particle ratios and the maximum values of the ratios in the case $\mu_s$ does not become negative ($\mu_s$=0).

---

[1]The maximally extended HG phase limit is defined for $T_0(\mu_q=0) \sim 183$ MeV, which is the maximum temperature for non-negative $\mu_s$ in the HG domain. Recent lattice QCD calculations give $T_0 \sim 175$ MeV for 2+1 quark flavours [7].

Note that negative $\mu_s$ means the particle yield ratio $\Omega^+/\Omega^- = \exp(-6\mu_s/T)$ becomes larger than unity. We find, within the error of the extrapolation, that at about $\sqrt{s}$ = 50 AGeV both ratios attain their largest values ~ (2.5 – 5) and ~ (1 – 1.5) for $\Omega^+/\Omega^-$ and $\Xi^+/\Xi^-$, respectively, compared to the maximum value of 1 and 0.55 for the case $\mu_s$ = 0, respectively.

## 7. Summary and Discussion

We have shown that the existence of an intermediate region of deconfined, massive and correlated quarks, in-between the hadronic and ideal quark-gluon phases, is realistic and necessary to explain some first experimental observations and theoretical conjectures. We have constructed an empirical EoS for this DQM phase, in terms of the order parameters and the mass-scaled partition functions of the HG and QGP phases, from which a relation for the strange quark-chemical potential is obtained in terms of $\mu_q$ and T. This empirical EoS includes mass-scaled strange hadrons, modified by appropriate factors of the order parameters and approximating the effects of the progressive decrease of the interaction–correlation with increasing temperature, as well as a mass-scaled QGP term. In effect, it describes realistically the gross features of the dynamics and characteristics of the DQM phase. The EoS indicates that an unambiguous and concise characteristic observable of this phase is the large negative strange quark-chemical potential. Analysis of data from sulfur-induced interactions at 200 AGeV give clear indications that the strange quark-chemical potential does assume large, negative values beyond the HG phase, in the deconfined domain.

Negative chemical potential appears also in condensed matter systems. For example, in a transition between weekly coupled Cooper pairs, with $\mu > 0$ and the usual BCS superconducting gap $|\Delta_k|$, and the strongly coupled diatomic pairs, with $\mu < 0$ and the corresponding gap $(\Delta_k^2 + \mu^2)^{1/2}$, representing an insulating system [29]. The analogy with the baryon-dense nuclear matter case is rather the opposite. The positive strange quark-chemical potential ($\mu_s > 0$) corresponds to a colour insulator (hadron gas state), whereas $\mu_s < 0$ to a colour (super)conductor (deconfined parton state).

An important argument of this work is that the light and strange quark fugacities, $\lambda_q$, $\lambda_s$ and the temperature are attributed to the equilibrated primordial state, to which Eq's (1,10) refer. Only in this case one would expect for a state in the deconfined phase to observe negative strange quark-chemical potential and temperatures in excess of those corresponding to deconfinement. This appears at first as 'impossible', since hadronization always takes place on the deconfinement-hadronization line, separating the HG phase from the DQM one and, therefore, these quantities should attain values only on this line. That is, always $\mu_s = 0$, and for the sulfur-induced interactions, for example, which have $\mu_q \sim 80$ MeV, a temperature T $\sim 176$ MeV [18,19].

To overcome this apparent difficulty, we propose that the conservation of fugacities $\lambda_i$ (i=u,d,s), is a characteristic property of strong interactions and thermodynamic equilibration in general, affecting thermally and chemically equilibrated states throughout the phase diagramme. That is, the quark number densities $n_i$ (i=u,d,s) and hence $\lambda_i$, since $n_i \sim \lambda_i$, once fixed in a chemically equilibrated primordial state, are constants of the entire sequent evolution process. This is contingent on an isentropic expansion and hadronization via a second order deconfinement phase transition (fast hadronization without mixed phase). The relation between $\lambda_i$ and T, and hence between $\mu_i$ and T in the primordial state[2] is defined by the strangeness neutrality equation, obtained from the partition function by imposing conservation of the S-quantum number.

The above statements have far-reaching consequences for defining and understanding the thermodynamic characteristics of the primordial state. They suggest that, if nuclear interactions form equilibrated states beyond the HG phase in the deconfined region with finite net baryon number density, the expansion to hadronization is isentropic and the confinement phase transition is of second order, one may determine the thermodynamic quantities of the state from the strange hadron yields, by employing an appropriate EoS. The observation of large negative strange quark-chemical potential in sulfur-induced interactions, which is not a characteristic of the hadronic phase, as well as temperatures in the range of 180 – 190 MeV (at $\mu_q \sim 80$ MeV), which are higher than the maximum temperature for deconfinement ($T_d \sim 176$ MeV at $\mu_q \sim 80$ MeV as given by SSBM), together with the proposed notion that negative strange quark-chemical potential indicates deconfinement, suggest that the thermodynamic quantities T, $\mu_i$ may indeed be deduced and attributed to the primordial state in the partonic phase.

If negative $\mu_s$-values, together with temperatures in excess of $T_d$ are confirmed at the proposed RHIC energies $20<\sqrt{s}<100$ AGeV, it will be a profound observation, indicating that negative strange quark-chemical potential is indeed a unique and well-defined signature of deconfinement, identifying the partonic phase. Of equal importance will be the possibility to determine the quantities $\mu_q$, $\mu_s$ and T, hence, the energy density and entropy of equilibrated primordial states situated *beyond* the hadronic phase, in the deconfined-quark region.

---

[2] The quantities $\mu_s$, $\mu_q$ and T are the Lagrange multipliers of strangeness, baryon number and energy, attaining their values in the equilibrated state.

On the basis of the present analysis we also conclude that the S+A interactions at 200 AGeV at the SPS are the only interactions to have produced an equilibrated partonic state beyond the hadronic phase. On the other hand, the thermodynamic parameters of the primordial state of the Pb+Pb interaction at 158 AGeV strongly indicate that its location is well within the hadronic phase.

**Appendix A.**

**1. Order Parameters**

In order to construct the EoS in the DQM region it is necessary to employ an analytic functional form of the order parameters. We have approximated the Wilson loop, obtained from lattice calculations, with two different functions of temperature:

$$R_d(T) = \frac{1}{1 + e^{-a(T-b)}} \tag{A.1}$$

and
$$R_d(T) = c\left(\frac{T - T_d}{T}\right)^\nu \tag{A.2}$$

In Eq.(A.1) a Fermi – type function is used, where the parameters {a, b} control the difference of the critical temperatures $T_d$ and $T_\chi$. They can be chosen arbitrarily, in order to obtain realistic compatibility with the lattice QCD results. In Eq.(A.2) the order parameter is described using a more common – in the theory of critical phenomena – function near the critical point $T_d$, where the critical exponent $v$ can be chosen according to universality class arguments or arbitrarily. The parameter $c$ simply controls the values of $R_d(T)$, so that asymptotically $R_d(T) \to 1$ as $T \to T_\chi$. Figures 6a,b show the approximated order parameter for different values of $\{a,b\}$ and of the critical exponent $v$, respectively. In Fig. 7 we show the two approximated order parameters, where we have taken $R_\chi(T) \equiv 1-R_d(T)$ and used the Fermi-type function, Eq.(A.1), with $\{a = 0.03, b = 268\}$.

**2.    Order parameters and the phase diagramme**

Since an empirical partition function for the DQM phase is used, Eq. (10), containing the parameters $R_d(T)$ and $R_\chi(T) \equiv 1-R_d(T)$, it is important to study the behaviour of the strange quark-chemical potential in the DQM region for the different approximations of $R_d(T)$. Figure 8 is a plot of $\mu_s$ vs. T, obtained for $\mu_q = 0.45T$ and the two different functions for $R_d(T)$. It is clear that the DQM region is characterized by negative strange quark-chemical potential independently of which $R_d(T)$ function is used in the EoS. Therefore, the change in the sign of the strange quark-chemical potential - from positive to negative - is a unique and well-defined indication of the quark-deconfinement phase transition and does not depend on the phenomenological parameters of the model. For only a qualitative comparison with data, we also show in the figure the two points of the sulphur-induced interactions. There is no adjustment of the parameters of the EoS to fit the data.

**3.    Order parameters and the effective quark masses**

The parameter $R_\chi(T)$ is included in the expressions of the effective quark and hadron masses. The use of temperature dependent quark masses is one of the essential aspects of the model, since it is a dynamical term in the equation of state. On the other hand, more precise theoretical models such as the N−J-L or the (non)linear sigma model, have also studied the temperature dependence of the light and strange quark masses. Figures 9a,b plot the light and strange quark masses as a function of temperature using the two different approximations of the order parameter, whereas Fig. 10 exhibits the same quantities but within the N−J-L model. The quark mass scaling used in our model is consistent with the results of an effective field theory model, despite the difference for $m_s^*$ which can be diminished by choosing appropriate $R_\chi(T)$ for strange quarks, evidently giving strong support to our approximations and results.

**Table 1.** Deduced values for T, $\mu_q$, $\mu_s$ from thermal – statistical model calculations and fits to experimental data for several nucleus-nucleus interactions. (*)Thermodynamic quantities extracted using several strange hadron yield ratios.

| | Interaction/Experiment | | | |
|---|---|---|---|---|
| | **Si + Au (14.6 AGeV)/E802** | | | |
| | Reference(*) | Reference [24] | Mean | |
| T (MeV) | 134 ± 6 | 135 ± 4 | 135 ± 3 | |
| $\mu_q$ (MeV) | 176 ± 12 | 194 ± 11 | 182 ± 5 | |
| $\mu_s$ (MeV) | 66 ± 10 | | 66 ± 10 | |
| | **Au + Au (11.6 AGeV)/E802** | | | |
| | Reference(*) | Reference [24] | Mean | |
| T (MeV) | 144 ± 12 | 121 ± 5 | 124 ± 5 | |
| $\mu_q$ (MeV) | 193 ± 17 | 186 ± 5 | 187 ± 5 | |
| $\mu_s$ (MeV) | 51 ± 14 | | 51 ± 14 | |
| | **Pb + Pb (158 AGeV)/NA49** | | | |
| | Reference(*) | Reference [24] | Reference [20] | Mean |
| T (MeV) | 146 ± 9 | 158 ± 3 | 156 ± 4 | 156 ± 3 |
| $\mu_q$ (MeV) | 74 ± 6 | 79 ± 4 | 78 ± 5 | 78 ± 3 |
| $\mu_s$ (MeV) | 22 ± 3 | | 26 ± 4 | 24 ± 2 |
| | **Pb + Pb (40 AGeV)/NA49** | | | |
| | Reference(*) | | | |
| T (MeV) | 147 ± 3 | | | |
| $\mu_q$ (MeV) | 136 ± 4 | | | |
| $\mu_s$ (MeV) | 35 ± 4 | | | |
| | **S + S (200 AGeV)/NA35** | | | |
| | Reference [21] | Reference [23] | Reference [22] | Mean |
| T (MeV) | 182 ± 9 | 181 ± 11 | 202 ± 13 | 188 ± 6 |
| $\mu_q$ (MeV) | 75 ± 6 | 73 ± 7 | 87 ± 7 | 78 ± 4 |
| $\mu_s$ (MeV) | – 60 ± 20 | – 58 ± 18 | | – 59 ± 13 |
| | **S + Ag (200 AGeV) NA35** | | | |
| | Reference [21] | Reference [23] | Reference [22] | Mean |
| T (MeV) | 180 ± 3 | 179 ± 8 | 185 ± 8 | 181 ± 4 |
| $\mu_q$ (MeV) | 79 ± 4 | 81 ± 6 | 81 ± 7 | 80 ± 3 |
| $\mu_s$ (MeV) | – 66 ± 20 | – 65 ± 23 | | – 65 ± 15 |

**Table 2.** The extrapolated values for the temperature T and the ratio $\mu_q/T$ obtained for the Au+Au interaction at RHIC, as well as predictions of the empirical EoS for $\mu_s/T$ and two strange-particle ratios.

| $\sqrt{s}$ (AGeV) | T (MeV) | $\mu_q / T$ | $\mu_s / T$ [a] | $\Omega^+/\Omega^-$ | $\Xi^+/\Xi^-$ |
|---|---|---|---|---|---|
| 20 | 181.76 ± 17.30 | 0.495 ± 0.016 | -0.03 ± 0.17 | 1.20 ± 1.22 | 0.42 ± 0.28 |
| 30 | 204.40 ± 19.35 | 0.418 ± 0.014 | -0.20 ± 0.10 | 3.24 ± 1.94 | 0.95 ± 0.38 |
| 40 | 220.45 ± 20.85 | 0.313 ± 0.012 | -0.21 ± 0.05 | 3.50 ± 1.05 | 1.23 ± 0.25 |
| 50 | 232.91 ± 22.02 | 0.299 ± 0.010 | -0.22 ± 0.05 | 3.77 ± 1.13 | 1.33 ± 0.27 |
| 60 | 243.10 ± 23.06 | 0.252 ± 0.008 | -0.19 ± 0.03 | 3.18 ± 0.57 | 1.30 ± 0.16 |
| 70 | 251.70 ± 23.81 | 0.213 ± 0.007 | -0.16 ± 0.01 | 2.68 ± 0.16 | 1.26 ± 0.05 |
| 80 | 259.15 ± 24.53 | 0.180 ± 0.006 | -0.14 ± 0.02 | 2.29 ± 0.19 | 1.21 ± 0.07 |
| 90 | 265.72 ± 24.16 | 0.152 ± 0.005 | -0.11 ± 0.02 | 1.99 ± 0.24 | 1.06 ± 0.04 |

[a] Calculated from the empirical DQM EoS

**Figure Captions**

1. Variation of the strange quark-chemical potential with temperature in the 3-region phase diagram. The curve $\mu_s(T)$ intersects the $(T,\mu_q)$-plane at the intersection point of the $\mu_q=0.45T$ line and the deconfinement line.

2. Variation of $\mu_s/T$ as a function of the temperature of the equilibrated primordial state. A deconfinement temperature of $T_d \sim 176$ MeV is assumed, whereas chiral symmetry is restored at $T_\chi \sim 3.5 T_d$.

3. Plot of the mean values of the temperature and strange quark-chemical potential, as obtained from different thermal model fits to experimental data.

4. Phase diagram with the SSBM deconfinement line and the location (T, $\mu_q$ average values) of several interactions, deduced from the analysis with thermal models.

5. Predicted particle yield ratios $\Omega^+/\Omega^-$ and $\Xi^+/\Xi^-$ for a finite baryon density region, obtained in Au + Au interactions at $\sqrt{s}$ = 20 - 90 AGeV. The lines correspond to the case $\mu_s=0$.

6 a,b. Approximated average thermal Wilson loop $R_d(T)$, obtained from Eq.(A.1) and Eq.(A.2) for different values of the model parameters {a, b} and *v*.

7. Plot of the approximated order parameters $R_d(T)$ and $R_\chi(T)=1-R_d(T)$ using Eq.(A.1) with a=0.03 and b=268.

8. Plot of the strange quark-chemical potential as a function of the temperature and the two different functions for $R_d(T)$, obtained for {a=0.03, b=268} and *v*=0.2, respectively.

9 a,b. The effective quark masses $m_q^*(T)$ and $m_s^*(T)$ calculated using the expressions of Eq.(A.1) and Eq.(A.2) for $R_\chi(T) = 1-R_d(T)$ and the values a=0.03, b=268 and *v*=0.2, c=1, respectively.

10. Results on the temperature dependence of the light and strange quark masses obtained from the N – J-L model [11].

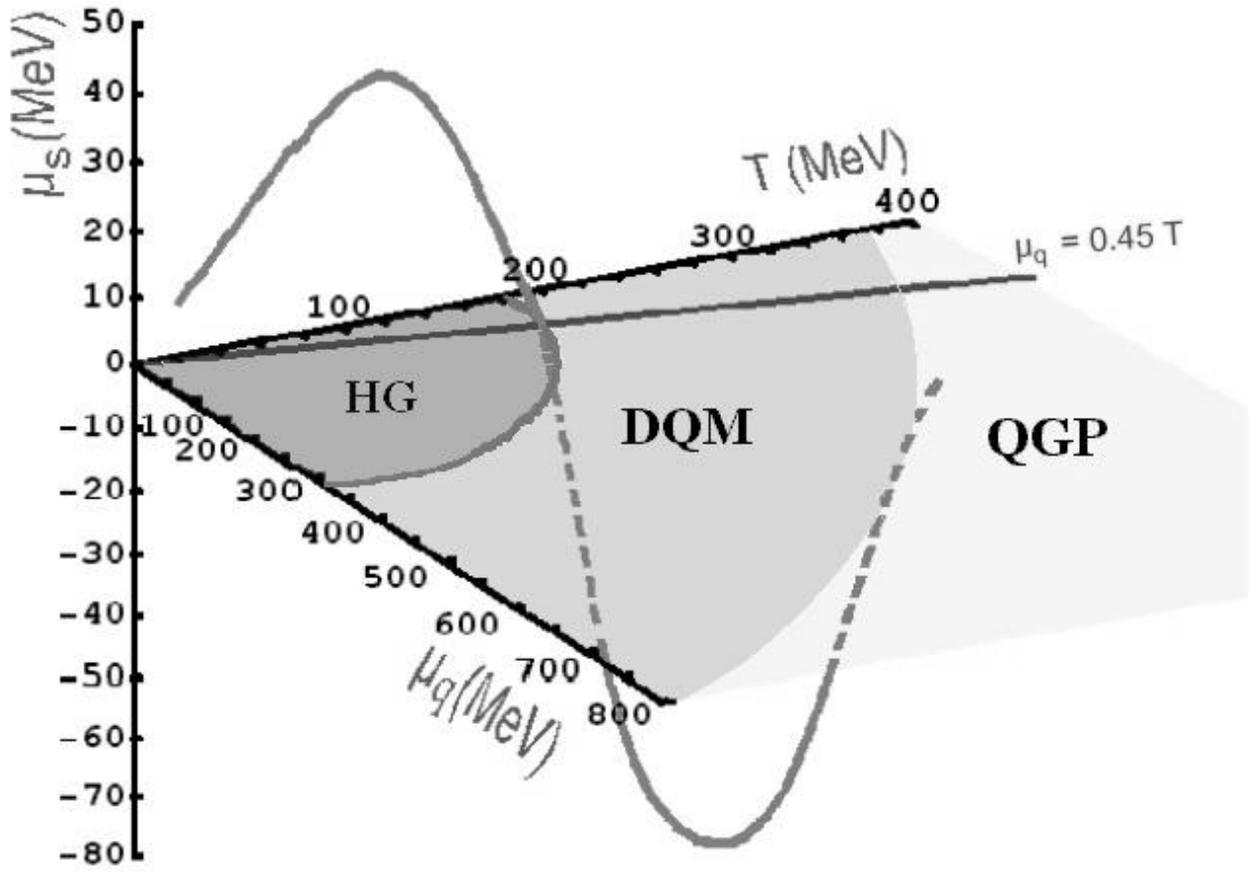

fig.1

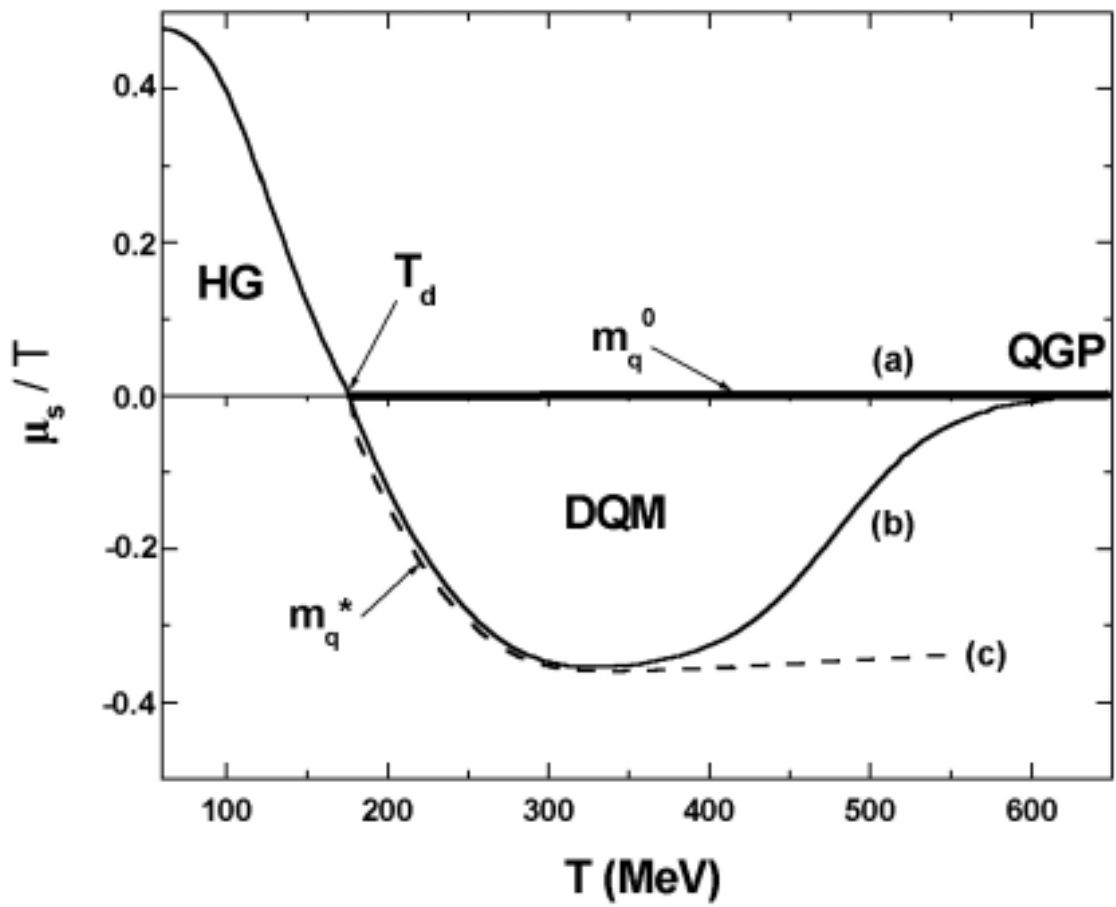

Figure 2

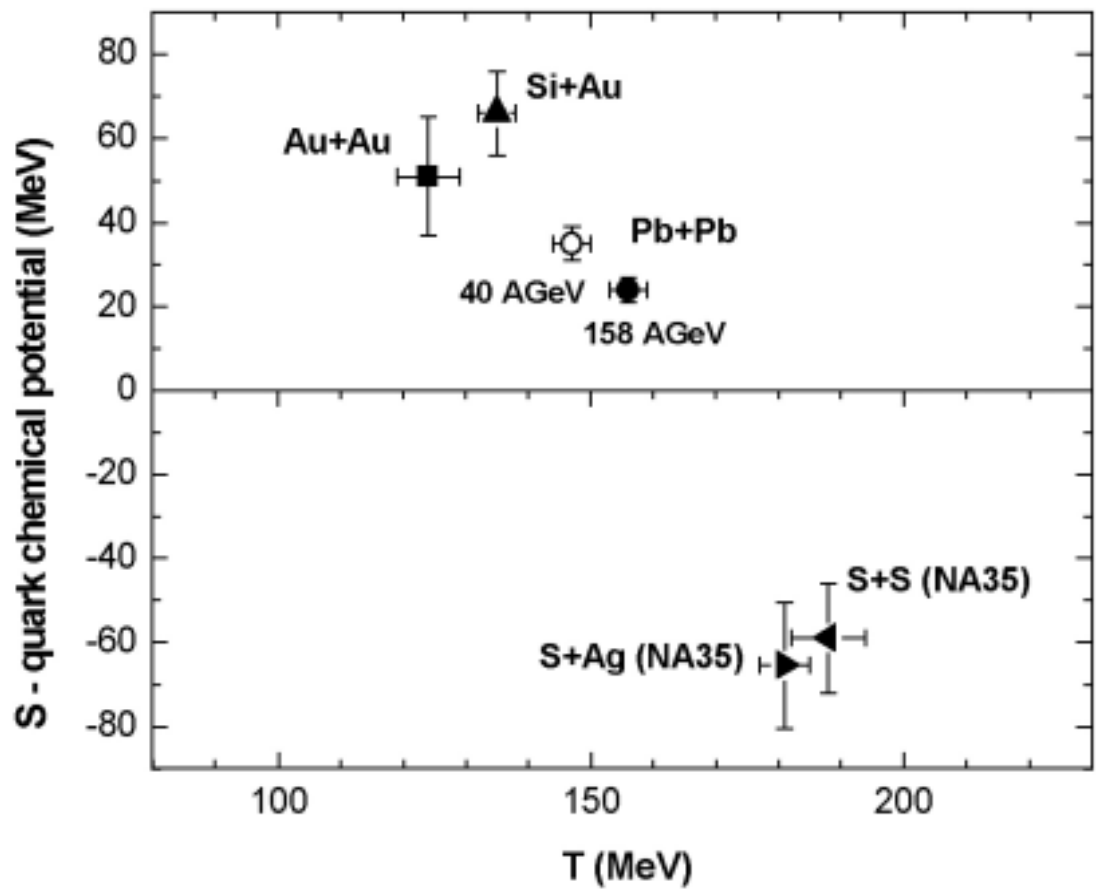

Figure 3

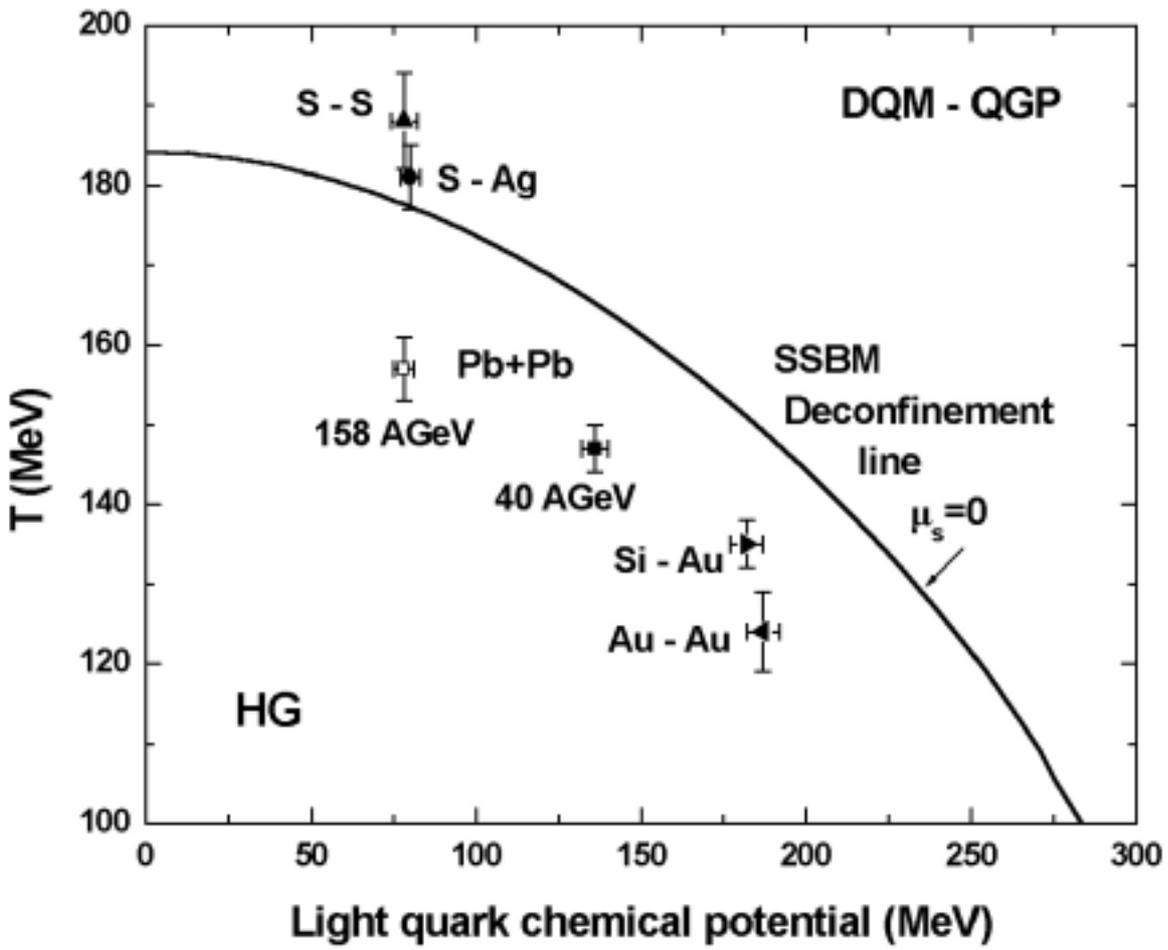

Figure 4

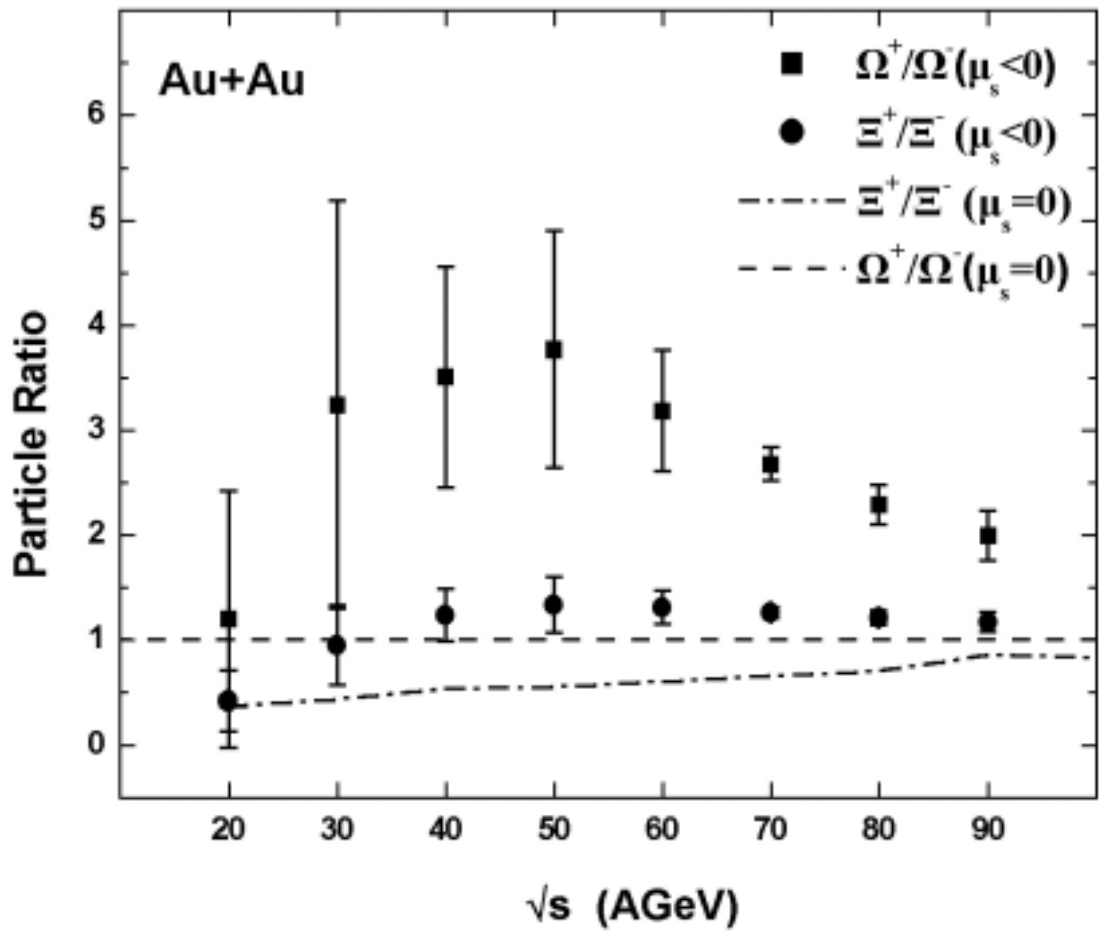

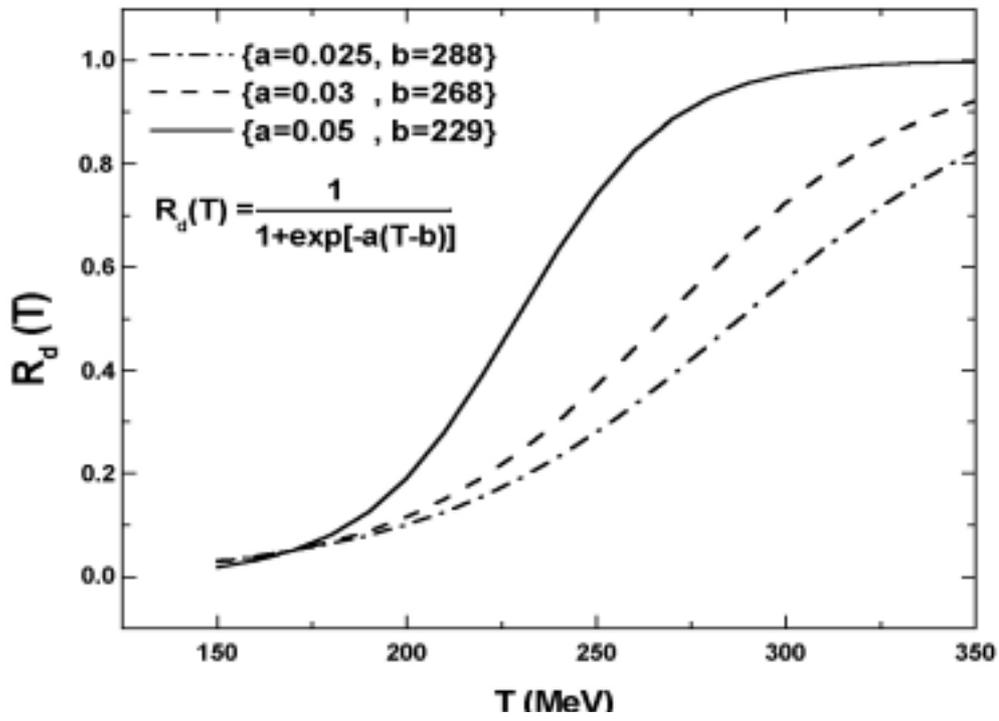

Figure 6a

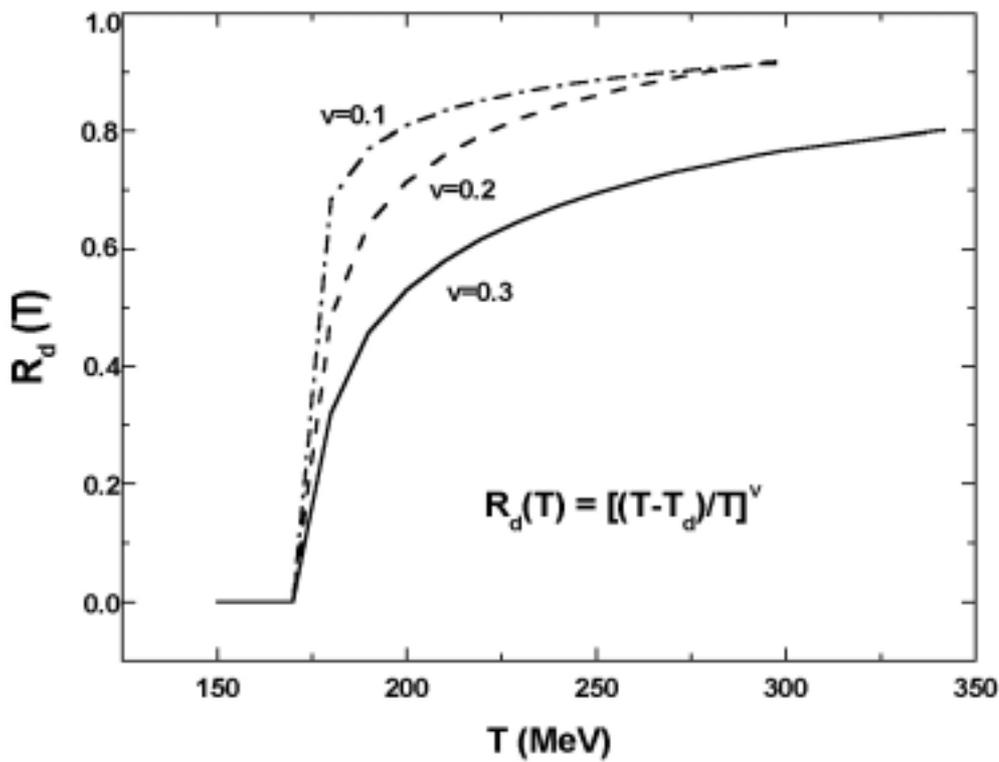

Figure 6b

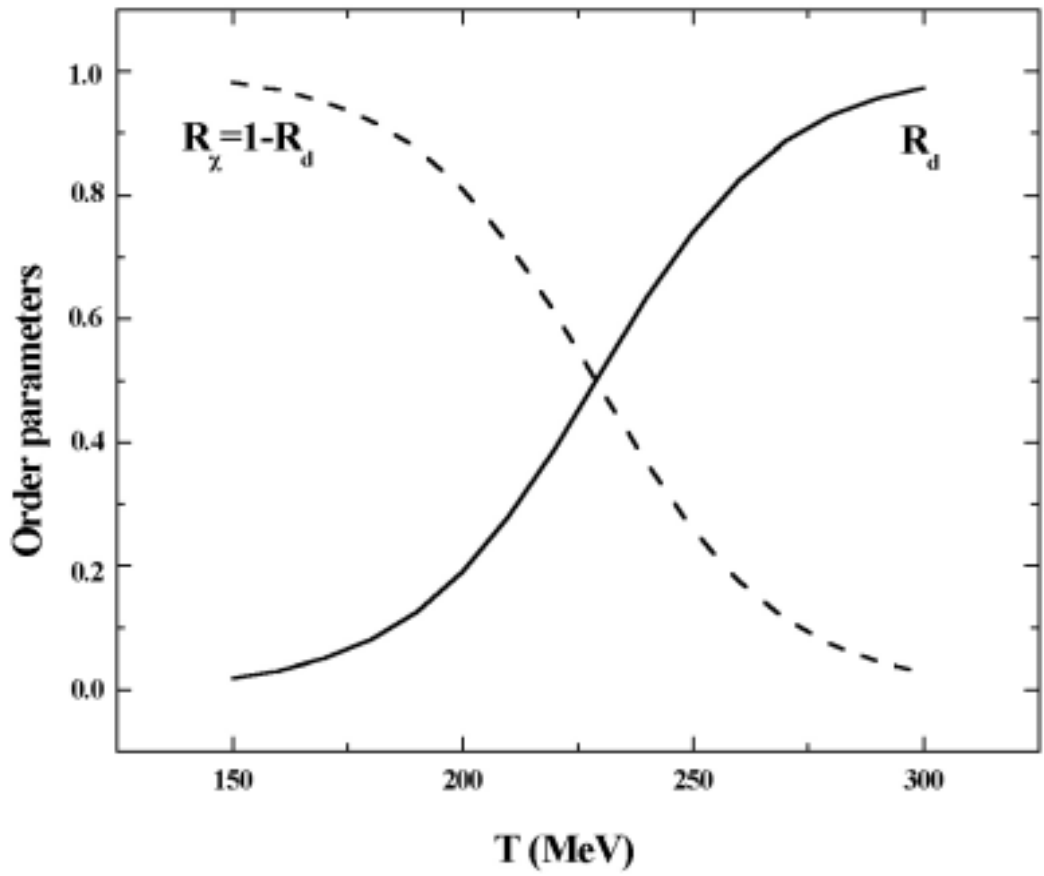

Figure 7

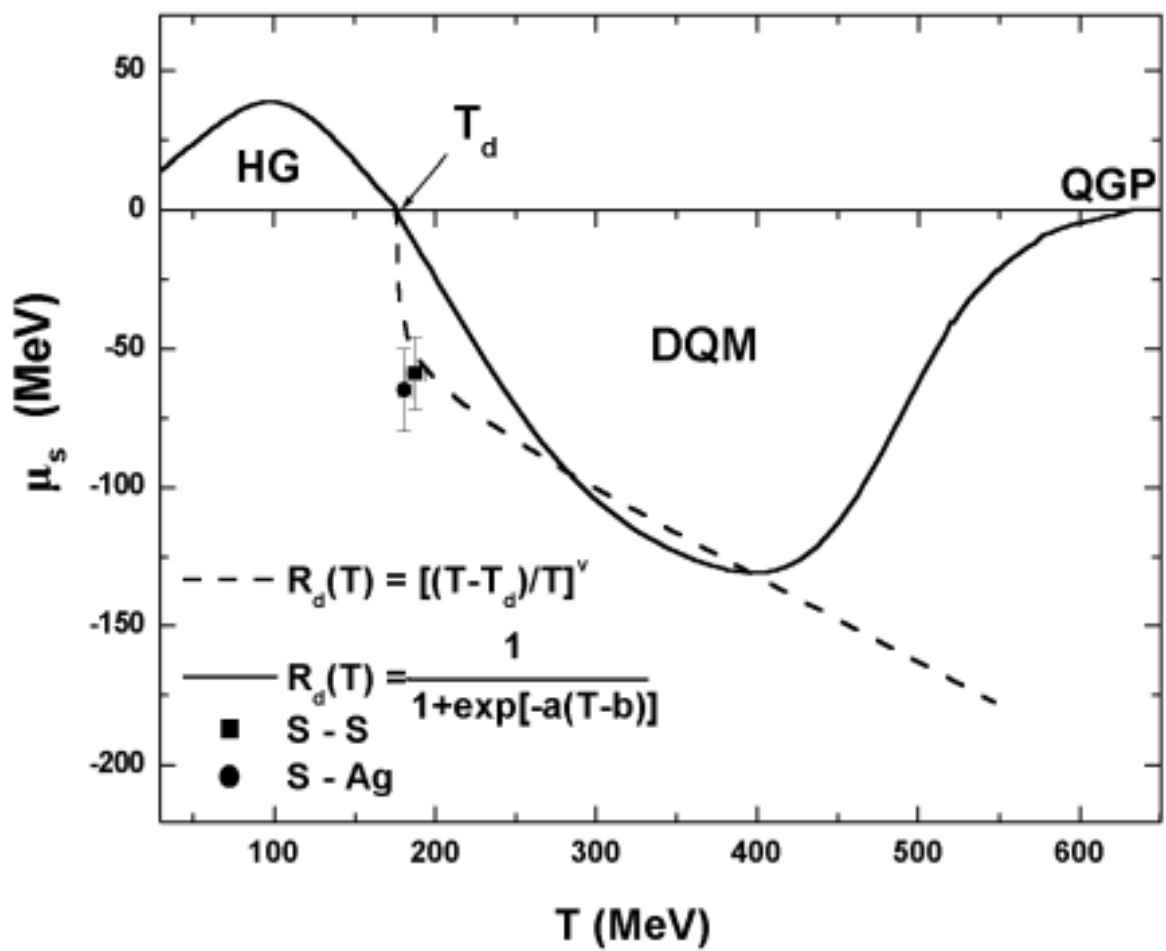

Figure 8

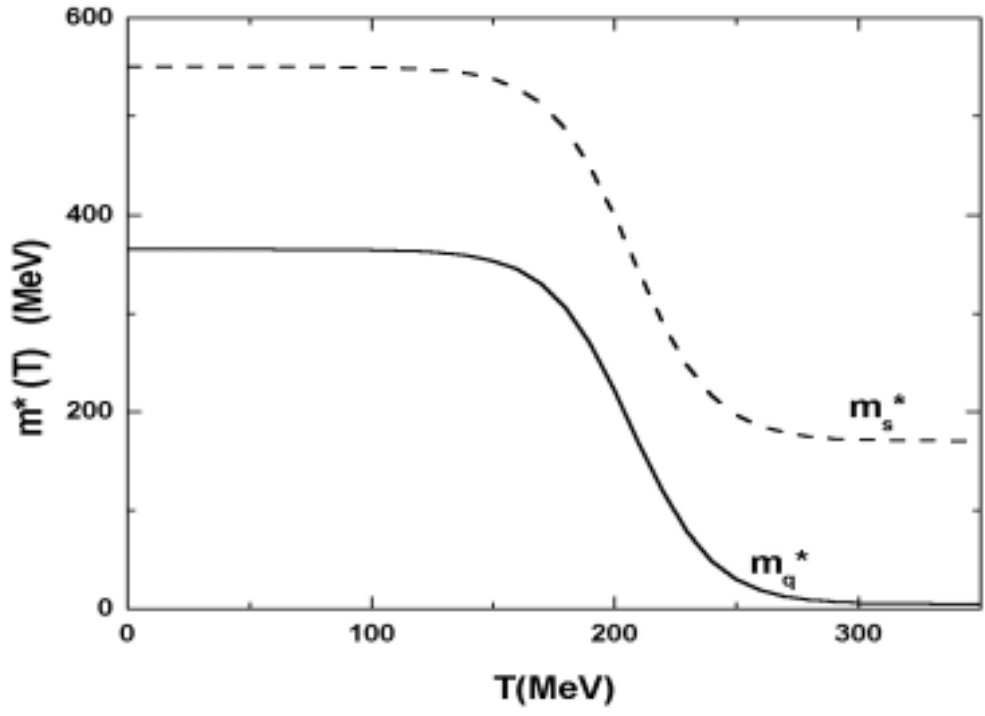

Figure 9a

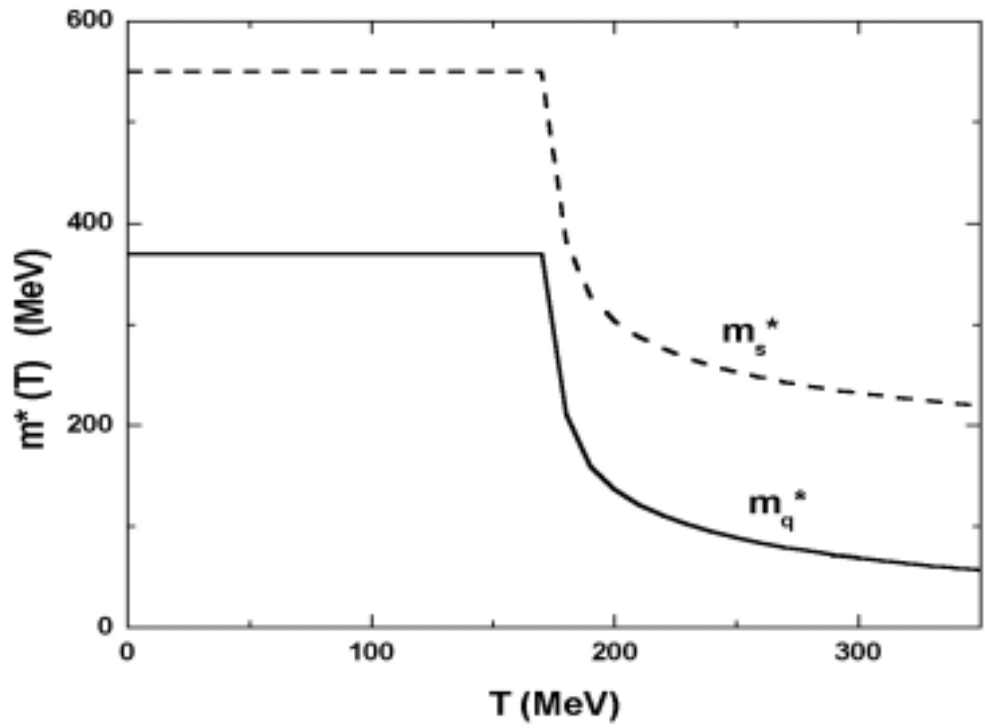

Figure 9b

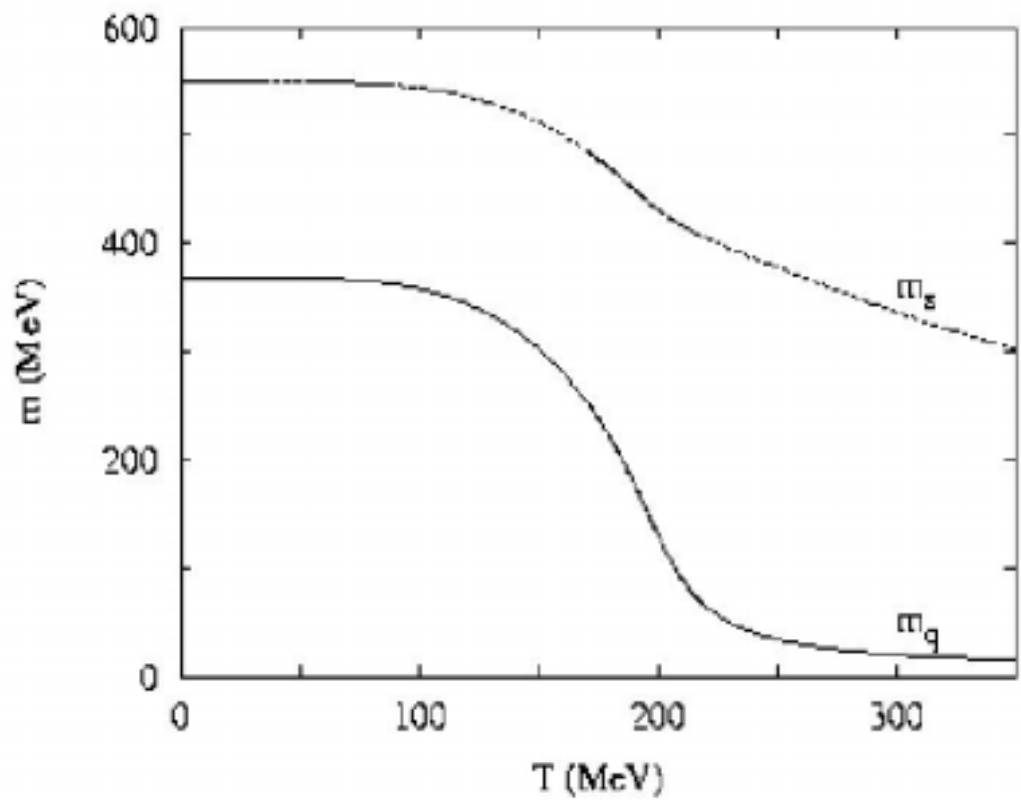

Figure 10